# Orientation effects on near-field radiative heat transfer between complex-shaped dielectric particles

Lindsay P. Walter[1] and Mathieu Francoeur[1,a)]

[1]Radiative Energy Transfer Lab, Department of Mechanical Engineering, University of Utah, Salt Lake City, UT 84112, USA

The effect of orientation on near-field radiative heat transfer between two complex-shaped superellipsoid particles of $SiO_2$ is presented. The particles under study are 50 nm in radius and of variable concavity. Orientation is characterized by the degree of rotational symmetry in the two-particle systems, and the radiative conductance is calculated using the discrete system Green's function approach to account for all electromagnetic interactions. Results reveal that the total conductance in some orientations can be up to twice that of other orientations when particles are at center-of-mass separation distances of 110 nm. Orientation effects are not significantly correlated with system rotational symmetries but are strongly correlated with the minimum vacuum gap distance between particles. As such, orientation effects on near-field radiative heat transfer are a consequence of particle topology, with more extreme topologies leading to a continuation of orientation effects at larger particle center-of-mass separation distances. The concave superellipsoid particles display significant orientation effects up to a center-of-mass separation distance approximately equal to 3.9 times the particle radius, while the convex superellipsoid particles display significant orientation effects up to a center-of-mass separation

[a)] Corresponding author. Email address: mfrancoeur@mech.utah.edu

distance approximately equal to 3.2 times the particle radius. In contrast to previous anisotropic, spheroidal dipole studies, these results of complex-shaped superellipsoid particles suggest that orientation effects become negligible when heat transfer is a volumetric process for all orientations. This work is essential for understanding radiative transport between particles that have non-regular geometries or that may have geometrical defects or abnormalities.

Micro/nanoparticles are gaining increasing interest for their potential to modulate radiative thermal transport[1–9]. In the near-field regime in which interparticle distances are less than the characteristic thermal wavelength defined by Wien's law, radiative transport between micro/nanoparticles is often dominated by shape-dependent surface polaritons and localized surface modes[10–12], opening the possibility for tuning the spectrum of heat transfer and thus spectral radiative properties[13,14]. Spectral tuning of radiative transfer between micro/nanoparticles finds application in a wide range of fields, from thermophotovoltaics[15–17], to radiative cooling[18,19], metamaterial design[20] and thermal circuitry[2,21].

Prior research suggests that near-field radiative heat transfer (NFRHT) between anisotropic particles may be tuned by varying particle orientation[22–26]. For example, researchers have studied NFRHT between SiC nanoparticles modeled as prolate and oblate spheroidal dipoles and found that the heat flux[22,23] and the time evolution of temperature[24] may be modulated by varying the orientation of particles. While some orientations result in minimal heat flux, other orientations result in heat flux values over two orders of magnitude greater than that of spherical dipoles of equivalent volume and center-of-mass separation distance[22,23]. This large range in heat flux has been attributed to the anisotropic polarizabilities of prolate/oblate spheroidal dipoles[22], and these results have led to the conclusion that anisotropic nanoparticles may be applied as a sort of thermal switch[22]. In dipole models such as these[22–24,26], the simplifying approximations require that the center-of-mass separation distance $d_c$ be greater than the particle characteristic length $L_{ch}$[27]. However, the numerical value of separation distances used in Ref. 22 was not reported, while the dipole approximation was applied to closely spaced particles that did not satisfy the required condition $d_c \gg L_{ch}$ in Ref. 23. Consequently, it is unclear whether these findings relating particle orientation to large differences in heat flux represent physical systems. Even when dipole approximations are applied within their regime of validity, these approximations are still quite restrictive and cannot capture the full near-field electromagnetic interactions required for design of real-world particle devices, such as Mie resonance-based metamaterials[28–30] and packed particle beds[1]. Except for a few studies involving two gold cylinders[31,32] and two graphene dimers[25], the impact of orientation on NFRHT between particles beyond the dipole approximation has not been systematically analyzed.

In this letter, we study the effect of orientation on NFRHT between two complex-shaped dielectric particles made of $SiO_2$, a material supporting surface phonon-polaritons (SPhPs) in the infrared. The particles are embedded in vacuum. NFRHT between these particles is modeled via the discrete system Green's function (DSGF) method[33], which is a numerically exact volume integral approach based on fluctuational electrodynamics[34]. In the DSGF method, all electromagnetic interactions are defined by computing a complete system Green's function. As such, the particle models presented here include multiple reflections and all higher order poles, thus enabling NFRHT predictions for vacuum gap distances down to the limit of applicability of fluctuational electrodynamics (~10 nm[35–37]) regardless of the particle characteristic length. Specifically, in order to determine the underlying physics driving orientation effects on NFRHT, we analyze the degree to which rotational symmetries of two-particle systems may be correlated with the radiative conductance. This work fills the knowledge gaps left from previous studies on particle orientation and is essential for providing the fundamental insight necessary to tune thermal radiation with particles both in the near and far field.

The complex-shaped particles studied in this work are modeled as superellipsoids of variable concavity. Superellipsoid particles were chosen because of their many-fold enhancement of photo/electrocatalytic behavior as compared with cubic nanoparticles[38,39], because they can easily be packed into a variety of superstructure formations[40,41], and because the rotational symmetries in two-particle systems may be systematically controlled. We model two different superellipsoid particle geometries: concave superellipsoids and convex superellipsoids [Figs. 1(a)–1(b)]. Superellipsoid surfaces are defined as

$$\left(\frac{x}{R_x}\right)^{p_x} + \left(\frac{y}{R_y}\right)^{p_y} + \left(\frac{z}{R_z}\right)^{p_z} = 1, \tag{1}$$

where $x$, $y$, and $z$ are Cartesian coordinates; $p_x$, $p_y$, and $p_z$ specify the curvature; and $R_x$, $R_y$, and $R_z$ are the axial radii. Here, $R_x = R_y = R_z = R = 50$ nm and $p_x = p_y = p_z = p$, where $p = 0.75$ defines concave superellipsoids and $p = 1.5$ defines convex superellipsoids. The characteristic length of each particle is taken as the largest radius and is the same for both superellipsoid shapes, $L_{ch} = R = 50$ nm.

The center-of-mass separation distances modeled here span 110 nm $\leq d_c \leq$ 300 nm, a range from the dipole limit down to distances at which vacuum gaps $d$ are only a fraction of the characteristic length of the particles. The DSGF method is applicable in this entire range of separation distances as long as adequately refined discretization is used[33] (more detail on particle discretization is provided in the supplementary material, Table SI and Fig. S1). Six different particle orientations were chosen for variable degree of rotational symmetries of the two-particle system, where orientations 1 and 2 have the highest degree of rotational symmetry and orientation 6 has the lowest degree of rotational symmetry [Fig. 1(c)]. The degree of rotational symmetry $n$ is defined as the number of isometries around all axes of symmetry in each two-particle system. Rotation angles and axes of symmetry for each isometry were found using the algorithm in Ref. 42 (see supplementary material, Sec. S2).

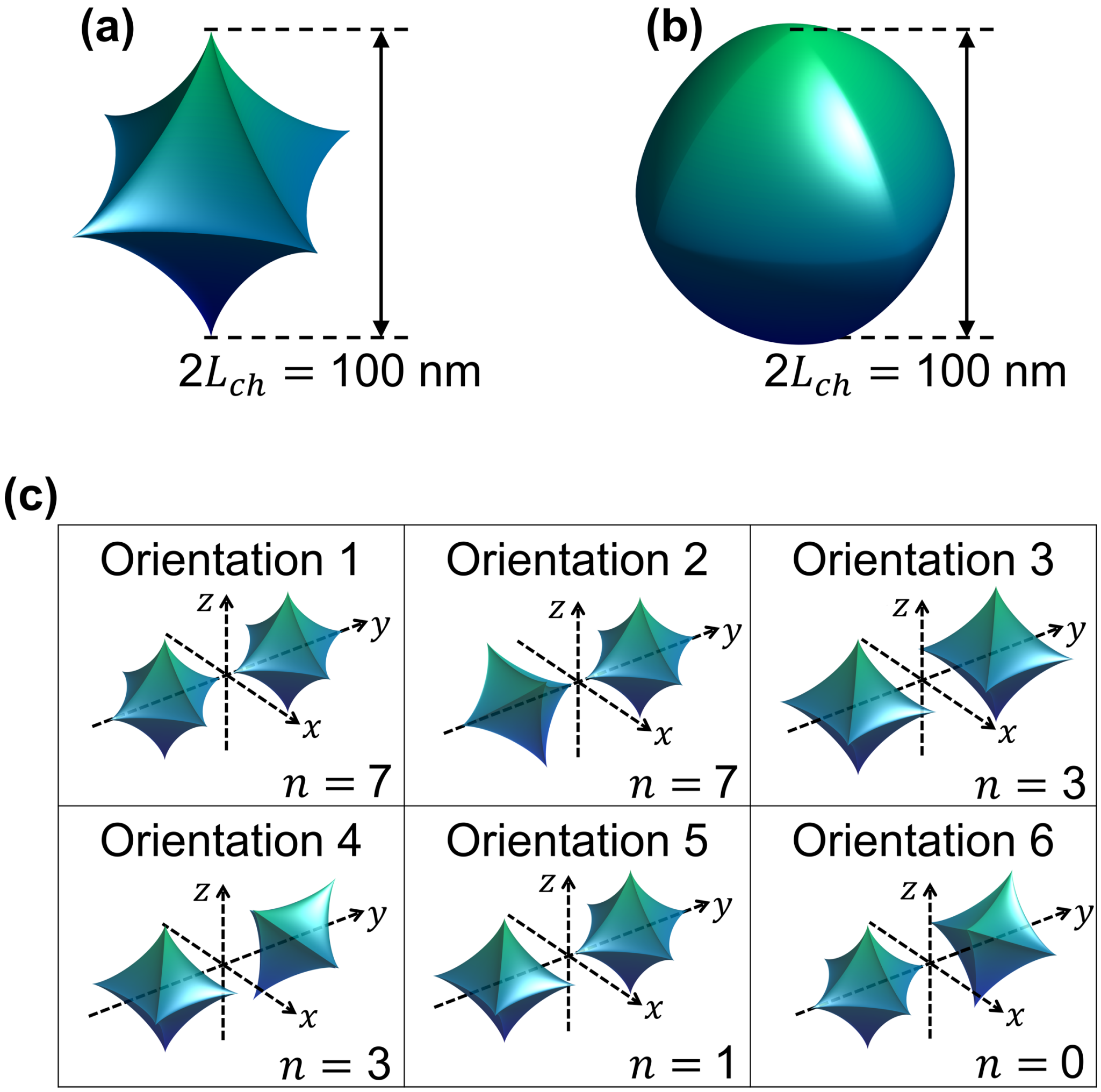

FIG. 1. Superellipsoid $SiO_2$ particles of variable shape and same characteristic length, $L_{ch} = 50$ nm: (a) concave superellipsoid ($p = 0.75$) and (b) convex superellipsoid ($p = 1.5$). (c) Two-particle systems of variable degree of rotational symmetry $n$ as calculated by the algorithm in Ref. 42. Particles are modeled using the software in Ref. 43.

Since one of the aims of this work is to determine whether orientation effects on NFRHT may be correlated with the degree of rotational symmetry in a two-particle system, the total and spectral conductance are chosen as the parameters for characterizing NFRHT. All conductance values are calculated at room temperature, $T = 300$ K. The total conductance $G_{t,AB}(T)$ and spectral conductance $G_{AB}(\omega, T)$ between particles $A$ and $B$ are defined as

$$G_{t,AB}(T) = \frac{1}{2\pi}\int_0^\infty G_{AB}(\omega,T)\,d\omega = \frac{1}{2\pi}\int_0^\infty \left[\frac{\partial\Theta(\omega,T')}{\partial T}\right]_{T'=T} \mathcal{T}_{AB}(\omega)\,d\omega, \tag{2}$$

where $\Theta(\omega,T)$ is the energy of a harmonic oscillator defined as $\Theta(\omega,T) = \hbar\omega\left[\exp\left(\frac{\hbar\omega}{k_B T}\right) - 1\right]^{-1}$, with $\hbar$ denoting Planck's constant, $\omega$ the angular frequency, $k_B$ the Boltzmann constant, and $T$ the temperature. $\mathcal{T}_{AB}(\omega)$ is the transmission coefficient between particles $A$ and $B$ defined as

$$\mathcal{T}_{AB}(\omega) = \sum_{i\in V_A}\sum_{j\in V_B} 4k_0^4 \Delta V_i \Delta V_j \mathrm{Im}[\varepsilon(\mathbf{r}_i,\omega)]\mathrm{Im}[\varepsilon(\mathbf{r}_j,\omega)]\mathrm{Tr}\left[\overline{\overline{\mathbf{G}}}(\mathbf{r}_i,\mathbf{r}_j,\omega)\overline{\overline{\mathbf{G}}}^{\dagger}(\mathbf{r}_i,\mathbf{r}_j,\omega)\right]. \tag{3}$$

In Eq. (3), $i$ and $j$ are indices for each cubic subvolume of the discretized lattice used in the DSGF method, sums are taken over the entire volume of particles $A$ and $B$, $k_0$ is the vacuum wavevector defined as $k_0 = \omega\sqrt{\mu_0\varepsilon_0}$, with $\mu_0$ and $\varepsilon_0$ the vacuum permeability and permittivity, respectively, $\Delta V_i$ is the volume of the $i$th subvolume used in the particle discretization, $\varepsilon$ is the dielectric function, $\mathbf{r}_i$ and $\mathbf{r}_j$ are the locations of the center points of the $i$th and $j$th subvolumes, $\overline{\overline{\mathbf{G}}}(\mathbf{r}_i,\mathbf{r}_j,\omega)$ is the $3\times 3$ DSGF tensor describing the interaction between subvolumes located at $\mathbf{r}_i$ and $\mathbf{r}_j$, and $\dagger$ denotes the conjugate transpose. The dielectric function of $SiO_2$ is calculated using a multiple-oscillator Lorentz model with parameters taken from Ref. 44,

$$\varepsilon(\omega) = \varepsilon_\infty + \sum_{n=1}^{3}\left[\frac{S_n}{1-\left(\frac{\omega}{\omega_{0,n}}\right)^2 - i\Gamma_n\left(\frac{\omega}{\omega_{0,n}}\right)}\right], \tag{4}$$

where $\varepsilon_\infty$ is the high-frequency permittivity limit, $S_n$ is the absorption strength, $\omega_{0,n}$ is the natural frequency, and $\Gamma_n$ is the damping constant.

The average runtime for a typical simulation of two particles at one orientation with 17002 total subvolumes was 44 core hours per frequency running on an AMD Rome processor with 64 cores and 256 GB of RAM. This runtime translates to 4930 total core hours, or a total of 3.2 days of real time, to calculate the integrated total conductance over 111 frequencies at one orientation and a single separation distance.

The total conductance plotted as a function of the center-of-mass separation distance reveals that there is no strong correlation between the degree of rotational symmetry in these two-particle systems and NFRHT trends [Figs. 2(a)–2(b)]. While particle orientation has a clear impact on the total conductance when particles are at the closest center-of-mass separation distance $d_c =$ 110 nm, this effect is not significantly related to the degree of system rotational symmetries. As can be seen in Figs. 2(a)–2(b), the high-symmetry systems (orientations 1 and 2) display the largest total conductance values at the closest center-of-mass separation distance $d_c = 110$ nm, but this relation does not hold for decreasing rotational symmetries. For example, orientation 6 has the lowest degree of rotational symmetry, but the total conductance of this orientation is only the third smallest out of all the orientations at the closest separation distance $d_c = 110$ nm [see insets in Figs. 2(a)–2(b)].

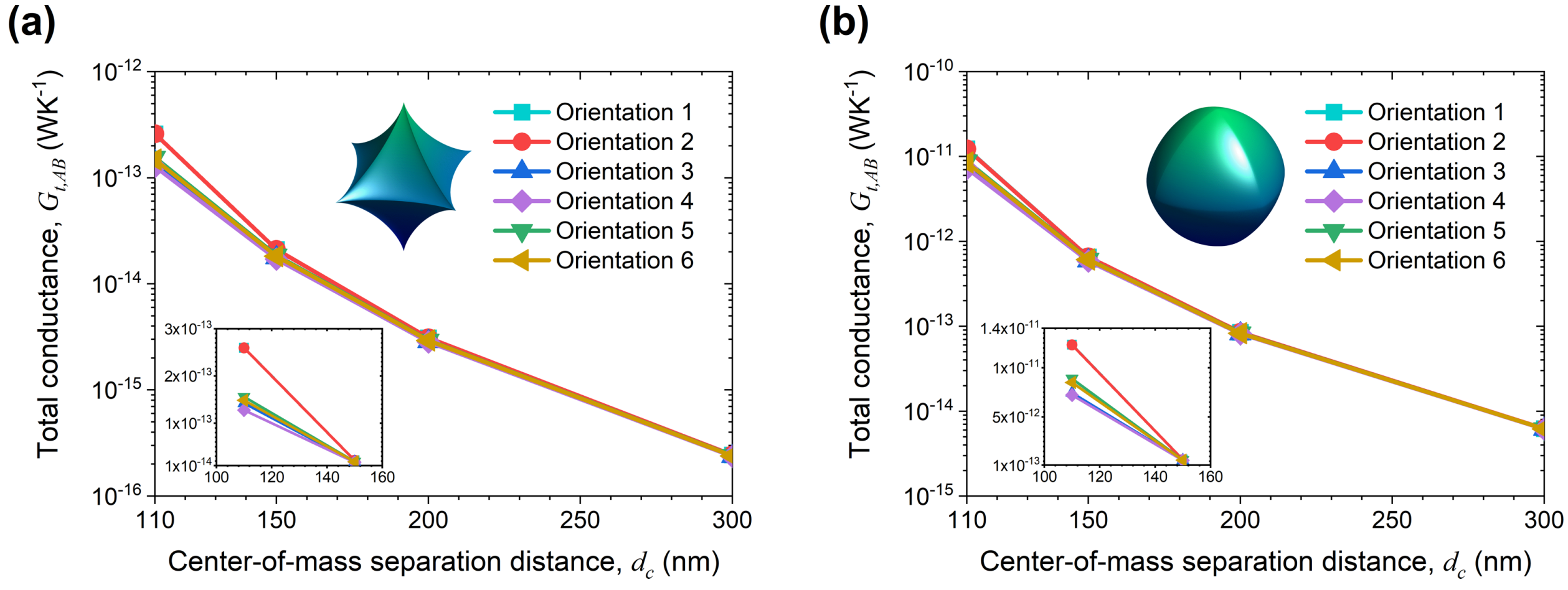

FIG. 2. Total conductance for (a) concave superellipsoid and (b) convex superellipsoid $SiO_2$ particles versus center-of-mass separation distance for all six orientations. Conductance is calculated at temperature $T = 300$ K. (Insets) Total conductance for the two closest center-of-mass separation distances, $d_c = 110$ nm and $d_c = 150$ nm (linear scale).

To quantify the extent to which orientation influences the total conductance at each center-of-mass separation distance, we consider the normalized range of the total conductance [Fig. 3(a)].

By considering the normalized, rather than the unnormalized, range of the total conductance, the orientation-dependent spread in conductance values can be directly compared across all particle separation distances. The normalized range $\bar{R}$ of a general parameter $X$ at a given center-of-mass separation distance $d_c$ is defined as the range of that parameter over all orientations divided by the mean of that parameter over all orientations,

$$\bar{R}[X(d_c)] = \frac{\text{range}[X_{orient,1}(d_c), X_{orient,2}(d_c), \dots, X_{orient,N}(d_c)]}{\text{mean}[X_{orient,1}(d_c), X_{orient,2}(d_c), \dots, X_{orient,N}(d_c)]}, \quad (5)$$

where $N = 6$ is the sample size of orientations. More accurate normalized range values could be obtained from a larger sample size of orientations. However, even for the sample of six orientations considered here, we expect the general trends observed in the normalized range of the total conductance to hold for larger orientation sample sizes. At $d_c = 110$ nm, the normalized range of the total conductance is 0.72 and 0.54 for concave and convex superellipsoid particles, respectively. These ranges correspond to significant orientation effects on NFRHT. The total conductance of concave superellipsoid particles in orientation 1 (i.e., the orientation with the largest total conductance, $G_{t,AB} = 2.60 \times 10^{-13}$ W/K) is about 2 times higher than that of these particles in orientation 4 (i.e., the orientation with the smallest total conductance, $G_{t,AB} = 1.28 \times 10^{-13}$ W/K). This difference is 1.7 times for convex superellipsoid particles in orientation 1 versus orientation 4. From the normalized range of the total conductance $\bar{R}(G_{t,AB})$, we can define two regimes of NFRHT, one in which orientation effects are significant with $\bar{R}(G_{t,AB}) \geq 0.1$ and one in which orientation effects are considered negligible with $\bar{R}(G_{t,AB}) < 0.1$. The center-of-mass separation distance at which this transition occurs varies with particle shape [Fig. 3(a)]. Interpolating between points, the transition occurs for concave superellipsoid particles at $d_c \approx 197$ nm and for convex superellipsoid particles at $d_c \approx 159$ nm. Orientation effects remain significant for concave superellipsoid particles at farther center-of-mass separation distances than for convex superellipsoid particles.

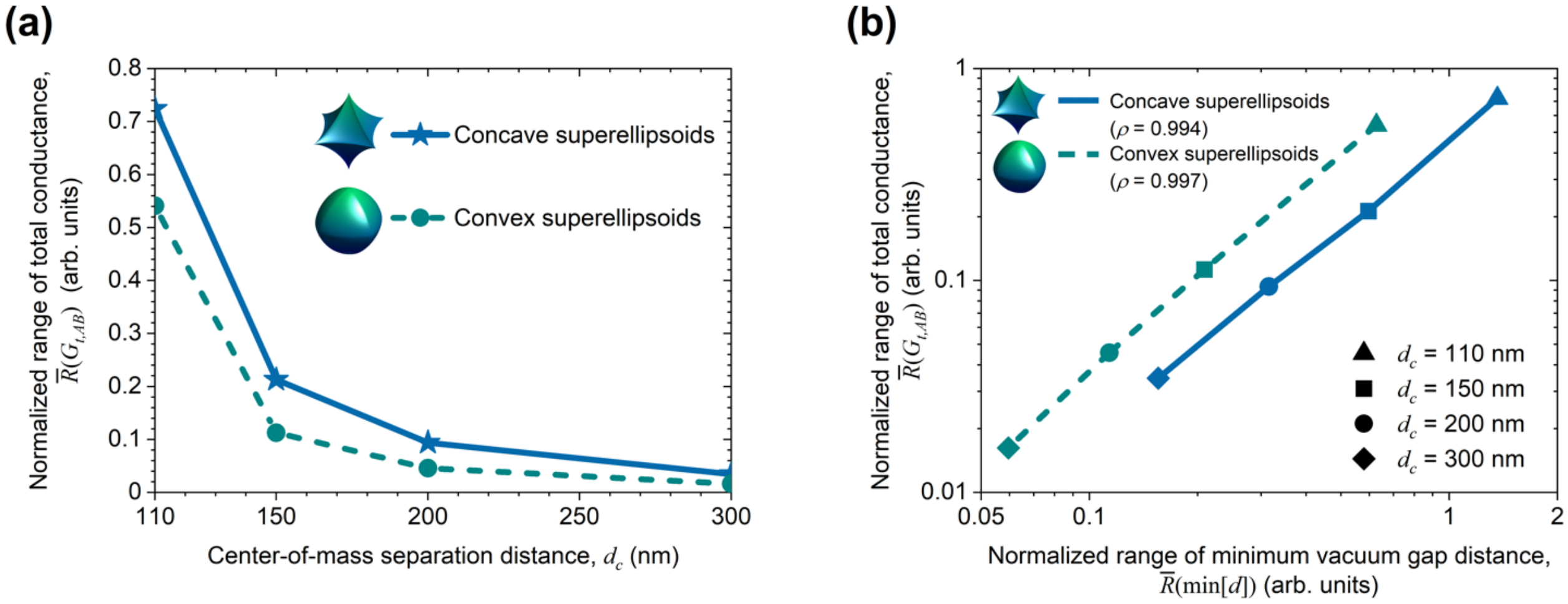

FIG. 3. (a) Normalized range of the total conductance (calculated at temperature $T = 300$ K) as a function of the center-of-mass separation distance for both particle shapes. (b) Correlation between the normalized range of total conductance and the normalized range of the minimum vacuum gap for both particle shapes evaluated at each center-of-mass separation distance. $\rho$ is the Spearman correlation coefficient calculated as $\rho = \frac{\sum_{i=1}^{N}(X_i-\mu_X)(Y_i-\mu_Y)}{\sqrt{\sum_{i=1}^{N}(X_i-\mu_X)^2}\sqrt{\sum_{i=1}^{N}(Y_i-\mu_Y)^2}}$, where $N = 6$ for the 6 different particle orientations, $X$ and $Y$ are the two parameters under analysis (here, the total conductance $G_{t,AB}$ and the minimum vacuum gap distance $\min[d]$), and $\mu_X$ is the mean of parameter $X$. Values of $\rho$ close to unity signify statistically significant correlation between parameters.

To identify the underlying factors for this difference in transition distances and to determine what exactly is driving orientation effects in the total conductance, we analyzed a variety of geometric parameters. From this analysis, only one parameter displayed similar trends as the total conductance data, namely, the minimum vacuum gap distance $d$ between particles (see supplementary material for minimum vacuum gap distance data, Fig. S3). Comparing the normalized range of the total conductance and the normalized range of the minimum vacuum gap at each center-of-mass separation distance, we see that these two parameters are highly correlated [Fig. 3(b) and supplementary material, Figs. S4(a)–S4(b)]. The Spearman correlation coefficient of these two normalized ranges is significant, with values of $\rho = 0.994$ for the concave superellipsoid particles and $\rho = 0.997$ for the convex superellipsoid particles. This similarity in trends between the minimum vacuum gap distance and the total conductance makes intuitive sense because heat transfer between $SiO_2$ particles in the near field is dominated by exponentially decaying SPhPs[45]. At the closest center-of-mass separation distance $d_c = 110$ nm, slight changes in the minimum vacuum gap distance with change in particle orientation result in large differences in the electromagnetic fields at interfacing surfaces and, therefore, in large differences in the total conductance. Confinement of heat transfer to the closest interfacing surfaces for particles at $d_c = 110$ nm is visualized by the spatial distribution of heat dissipation over the volume of each particle (see Fig. 4 for concave superellipsoid particles). At $d_c = 110$ nm, NFRHT is predominantly a surface, rather than a volumetric, phenomenon, and, consequently, particle orientation matters.

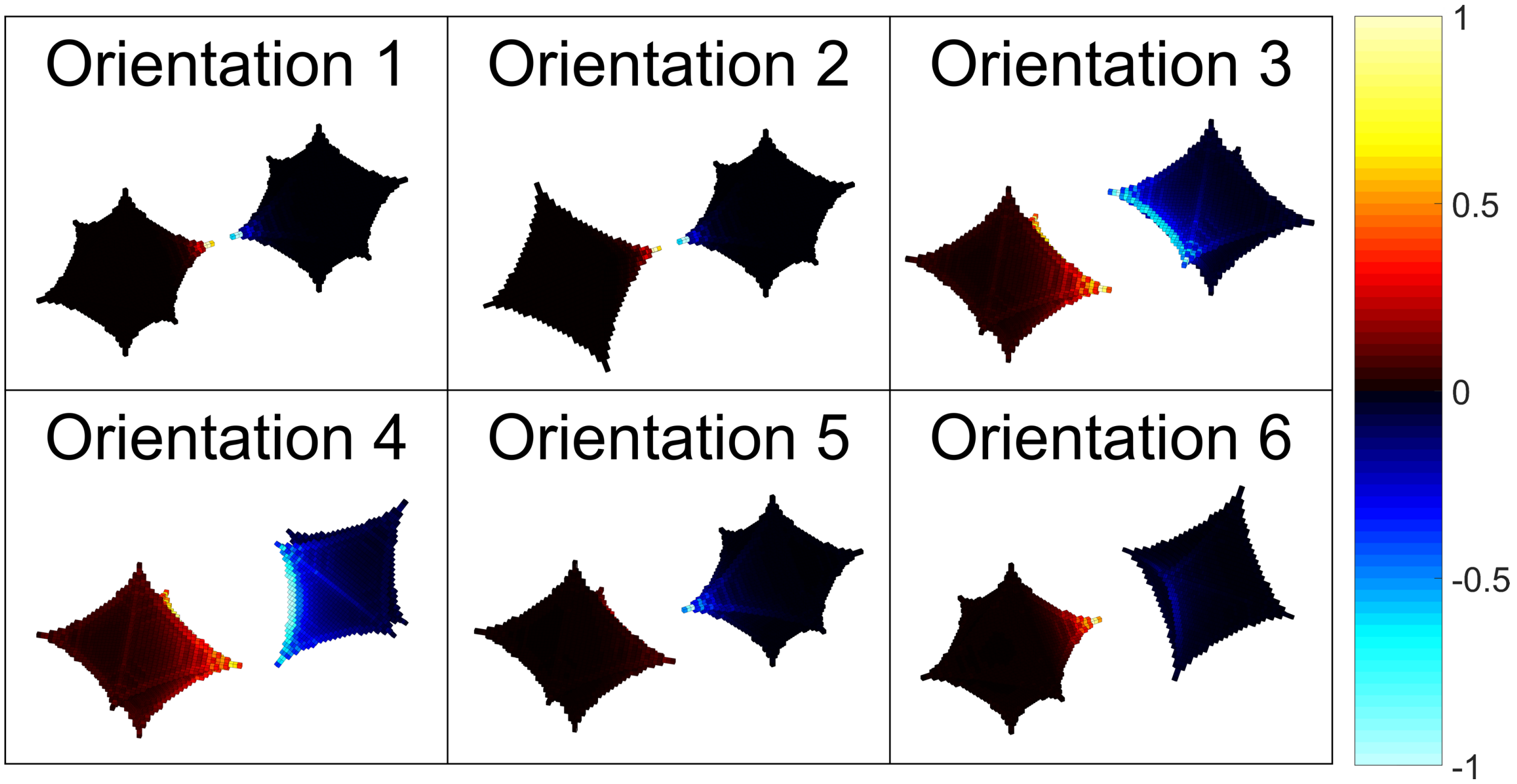

FIG. 4. Spatial distribution of heat dissipation for concave superellipsoid particles of variable orientation at center-of-mass separation distance $d_c = 110$ nm. The hot particle is at temperature $T = 300$ K, and the cold particle is at $T = 0$ K. Heat map colors are normalized by the highest heat dissipation value in each two-particle system, where positive values represent heating and negative values represent cooling. Units are arbitrary.

To see more clearly the effect of particle orientation on SPhP resonances, we analyze the spectral conductance of particles and compare the spectra against that of a perfect sphere of the same volume and center-of-mass separation distance [see Figs. 5(a)–5(d) for concave superellipsoid particles and the supplementary material for convex superellipsoid particles, Figs. S5(a)–S5(d)]. As shown in the literature on non-spherical particles of regular geometries[10,22,31,46] and in our previous work on irregularly shaped particles[33], the conductance spectra of non-spherical particles deviate from that of perfect spheres at all center-of-mass separation distances within the near-field regime. Spectral deviation for non-spherical particles is again seen here for superellipsoid particles of all shapes, particularly as it relates to shifts in SPhP resonances. At the closest center-of-mass separation distance $d_c = 110$ nm, the spectral conductance of particles is a function of particle orientation, especially near SPhP resonances [Fig. 5(a) and supplementary material, Fig. S5(a)]. This is expected from the previous total conductance results. At the largest center-of-mass separation distance $d_c = 300$ nm, the conductance spectra of different particle orientations converge to a single line, illustrating that NFRHT has become a volumetric process and that orientation effects are negligible [Fig. 5(d) and supplementary material, Fig. S5(d)].

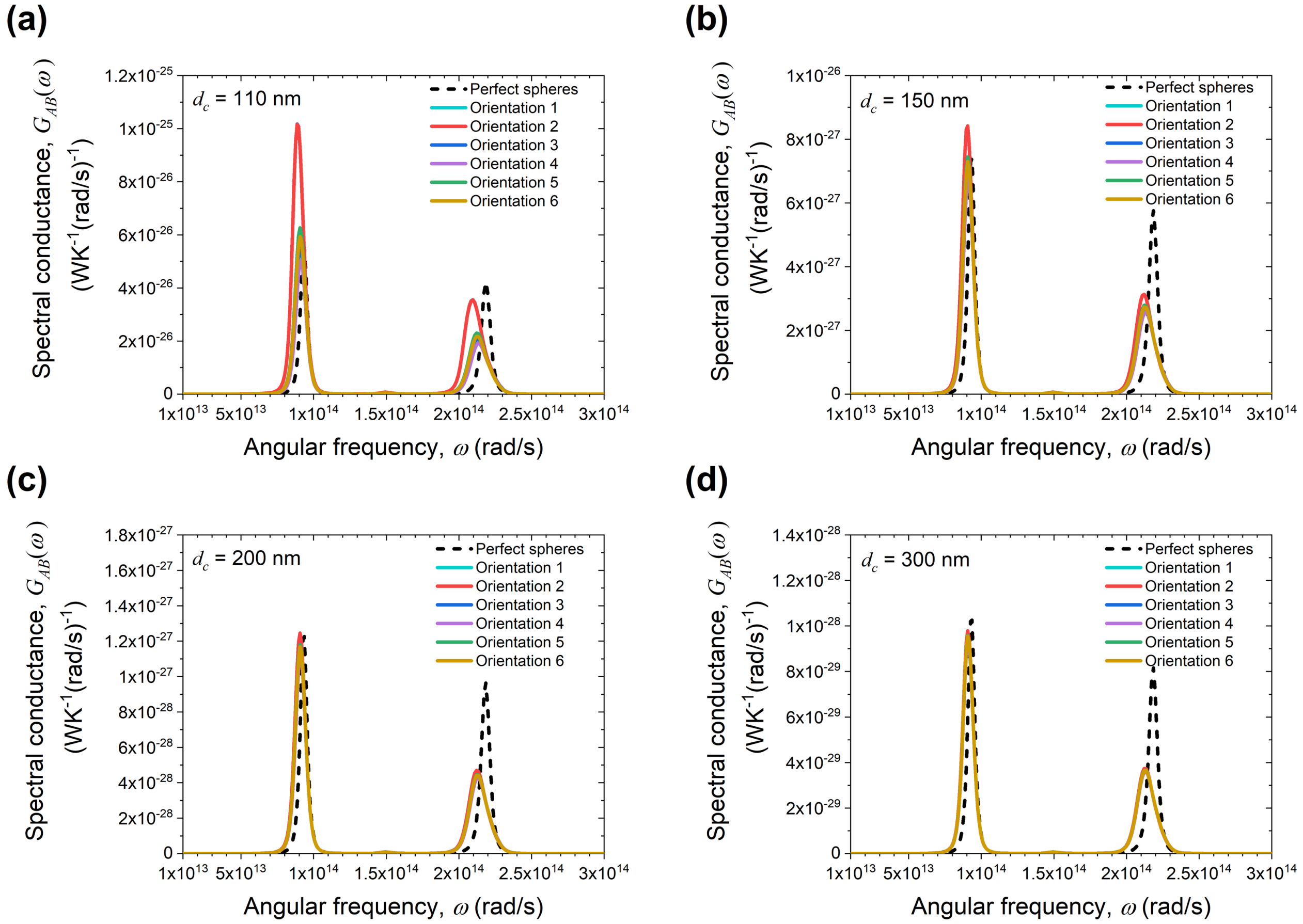

FIG. 5. Spectral conductance at temperature $T = 300$ K of concave superellipsoid $SiO_2$ particles at center-of-mass separation distances (a) $d_c = 110$ nm, (b) $d_c = 150$ nm, (c) $d_c = 200$ nm, and (d) $d_c = 300$ nm. The spectra of perfect spheres of equivalent volume and center-of-mass separation distance are calculated analytically using the method outlined in Ref. 44.

The dipole approximation in NFRHT is typically deemed valid when $d_c \gtrsim 3L_{ch}$[3,27,47]. When defining the regimes in which particle orientation is and is not significant to NFRHT, however, this relation between particle characteristic length and separation distance is insufficient because it does not account for the geometric shape and spread of particles. These factors are critical in determining precisely when orientation effects should and should not be considered. Using the aforementioned tolerance of 0.1 in the normalized range of the total conductance, orientation effects are considered negligible when $d_c \gtrsim 3.9L_{ch}$ for concave superellipsoid particles and $d_c \gtrsim 3.2L_{ch}$ for convex superellipsoid particles. In these relations, the multiplicative factor relating the center-of-mass separation distance $d_c$ and the particle characteristic length $L_{ch}$ is different for each particle shape. The relation may be written in the form $d_c \gtrsim CL_{ch}$, where $C$ is the shape-dependent multiplicative factor. Larger multiplicative factors $C$ reflect a greater spread in the topology of particles that leads to a greater range in vacuum gaps in two-particle systems of variable orientation. For example, we expect the factor $C$ to be larger than that of the superellipsoids shown here when the particles under study have high geometric anisotropy, such as long wires or flat disks. This prediction is consistent with NFRHT trends seen in clusters of nanoparticles of variable spread and orientation[48,49]. Therefore, we recommend that particle shape, especially the degree of topological variation, be taken into account when deciding whether full-scale simulations or simplified approximations are appropriate for modeling NFRHT in a given system of particles.

In conclusion, we have shown that the degree of rotational symmetries in systems of two $SiO_2$ superellipsoid particles does not appear to be significantly correlated with NFRHT trends. Instead, orientation effects on NFRHT between particles are strongly correlated ($\rho \approx 0.99$) with the minimum vacuum gap separation distance, and, thereby, a function of particle topology. This result relating orientation effects on NFRHT to the minimum vacuum gap between particles and to particle shape has important consequences in engineering micro/nanoparticle systems for radiative thermal management. Given a group of particles that may contain defects or irregularities, one may approximate when particle orientation effects will and will not be significant for NFRHT by analyzing a representative sample of geometric parameters. Additionally, this work suggests that orientation effects on NFRHT between complex-shaped particles become negligible once particles are in the volumetric regime of NFRHT, a result that differs from previous dipole approximation models[22–24]. A useful direction for future work is the analysis of orientation effects on NFRHT between particles of other shapes. Such analysis would aid in determining more general, predictive regimes maps of NFRHT so that appropriate modeling techniques may be employed for a given system of particles.

## SUPPLEMENTARY MATERIAL

See the supplementary material for the $SiO_2$ superellipsoid particle discretizations; rotational symmetries in the two-particle systems; comparison of the minimum vacuum gap distance between two superellipsoid $SiO_2$ particles for all orientations and center-of-mass separation distances; comparison of trends for the normalized range of the total conductance and the normalized range of the minimum vacuum gap distance for two superellipsoid $SiO_2$ particles; and the spectral conductance between two convex superellipsoid $SiO_2$ particles.

## ACKNOWLEDGEMENTS

This work was supported by the National Science Foundation (Grant No. CBET-1952210). L.P.W. acknowledges that this material is based upon work supported by the National Science Foundation Graduate Research Fellowship under Grant No. DGE-1747505. Any opinions, findings, and conclusions or recommendations expressed in this material are those of the authors and do not necessarily reflect the views of the National Science Foundation. The support and resources from the Center for High Performance Computing at the University of Utah are gratefully acknowledged.

## DATA AVAILABILITY

The data and code that support the findings of this study are available from the corresponding author upon reasonable request.

# Supplementary Material

Orientation effects on near-field radiative heat transfer between complex-shaped dielectric particles

Lindsay P. Walter[1] and Mathieu Francoeur[1,a)]

[1]Radiative Energy Transfer Lab, Department of Mechanical Engineering, University of Utah, Salt Lake City, UT 84112, USA

[a)] Corresponding author. Email address: mfrancoeur@mech.utah.edu

## S1. $SiO_2$ SUPERELLIPSOID PARTICLE DISCRETIZATIONS

The cubic lattice discretizations used in the discrete system Green's function[1] simulations are presented in Table SI and Fig. S1. Concave superellipsoid and convex superellipsoid particles are discretized along a cubic lattice.

TABLE SI. Number and size of subvolumes used in the discretization of each particle.

| Particle shape | Total number of subvolumes per particle | Length of a subvolume [nm] |
|---|---|---|
| Concave superellipsoid | 8,501 | 2.0408 |
| Convex superellipsoid | 7,231 | 3.7037 |

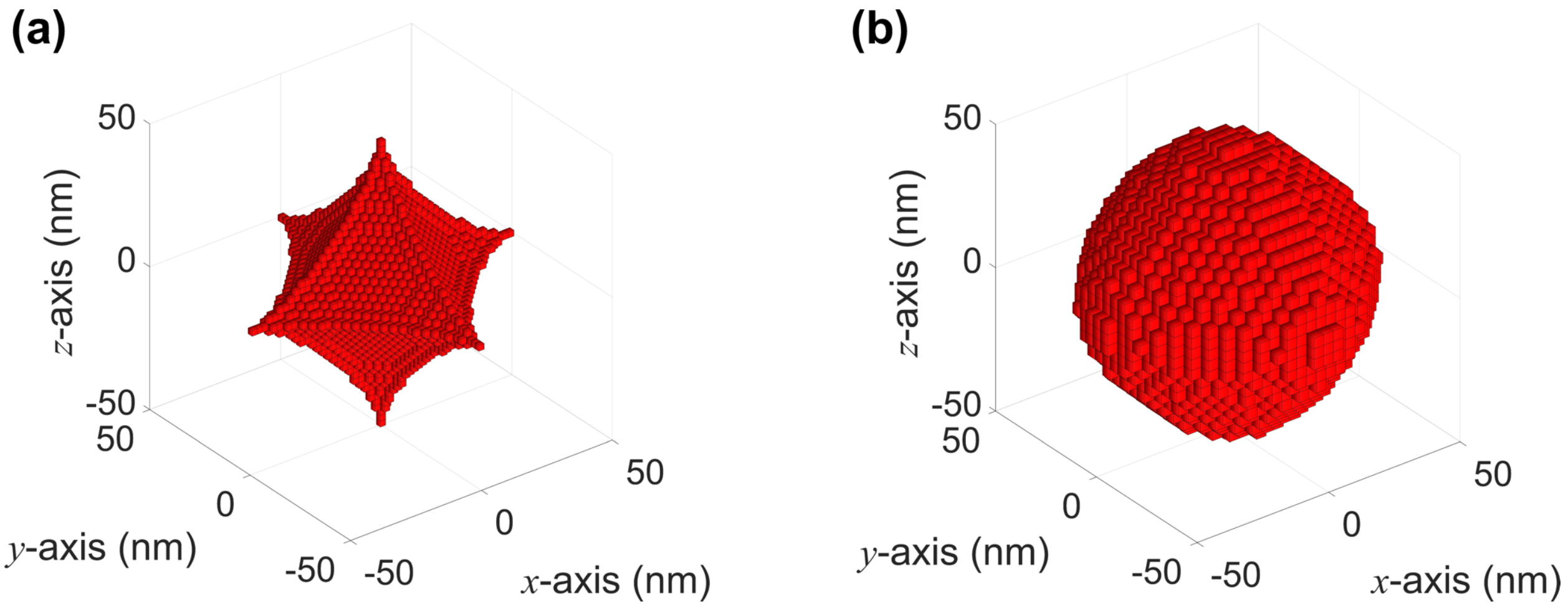

FIG. S1. Cubic lattice discretizations for (a) concave superellipsoid and (b) convex superellipsoid particles.

## S2. ROTATIONAL SYMMETRIES IN THE TWO-PARTICLE SYSTEMS

The degree of rotational symmetry $n$, defined as all two-particle rotational isometries, is found using the method in Ref. 2. In this method, the known rotational symmetries of a single superellipsoid particle are used as inputs in the following equation to find all possible two-particle isometries $K_{ij}$ in a complete, deterministic fashion:

$$K_{ij} = J_{ij} T_i^{-1} A T_i, \text{ with } A \in G_i, \tag{S1}$$

where subscripts $i$ and $j$ refer to each particle in the system, $J_{ij}$ is a known transformation seed mapping particle $i$ to particle $j$ in the global coordinate system, $T_i$ is the 4-by-4 affine transformation matrix defining the translation of the center of mass of particle $i$ to the center of mass of the two-particle system, and $A$ is a single isometry from the group of all isometries $G_i$ for the individual particle $i$ as written in the local coordinate system for particle $i$.

Once all possible two-particle isometries are found, the $K_{ij}$ transformation matrices are applied to each particle in the two-particle system to verify if they are a true match. A transformation matrix is considered a match, and a true system isometry is found, when

$$K_{11} S_1 = S_1 \text{ and } K_{22} S_2 = S_2, \tag{S2}$$

or

$$K_{12} S_1 = S_2 \text{ and } K_{21} S_2 = S_1, \tag{S3}$$

where $S_1$ and $S_2$ are the discretized surfaces of particle 1 and 2, respectively. To determine if two surfaces are equivalent, a symmetry measure similar to that used in Ref. 2 is defined,

$$d_M(S_i, S_j) = \max_{p \in V_{S_i}} \left( \min_{q \in V_{S_j}} \parallel p - q \parallel \right), \tag{S4}$$

where $V_{S_i}$ and $V_{S_j}$ are the vertex points of the discretized surfaces $S_i$ and $S_j$, respectively. Surfaces are considered equivalent when the symmetry measure $d_M$ is close to zero.

Once all true rotational isometries are found, the axis of rotation and angle of rotation for every isometry may be calculated from the $K_{ij}$ transformation matrices. The vector defining the axis of rotation $v$ is an eigenvector of the $K_{ij}$ matrix with eigenvalue $\lambda$ of 1,

$$K_{ij}v = \lambda v \text{ for } \lambda = 1. \tag{S5}$$

The absolute value of the angle of rotation $\alpha$ can then be found as

$$|\alpha| = \cos^{-1}\left[\frac{\mathrm{Tr}(K_{ij})-1}{2}\right]. \tag{S6}$$

All isometries for each two-particle system are defined by their axes of rotation and angles of rotation in Table SII. The final value of the degree of rotational symmetry $n$ is the number of all isometries, excluding the identity.

TABLE SII. Breakdown of rotational symmetries for each particle orientation.

| Orientation | Axis of rotation | Angle(s) of rotation | Degree of rotational symmetry, $n$ |
|---|---|---|---|
| 1 | $\begin{bmatrix}1\\0\\0\end{bmatrix}$ | $\pi$ | 7 |
| | $\begin{bmatrix}0\\1\\0\end{bmatrix}$ | $-\frac{\pi}{2}, \frac{\pi}{2}, \pi$ | |
| | $\begin{bmatrix}0\\0\\1\end{bmatrix}$ | $\pi$ | |
| | $\begin{bmatrix}\sqrt{2}/2\\0\\\sqrt{2}/2\end{bmatrix} = \begin{bmatrix}0.70711\\0\\0.70711\end{bmatrix}$ | $\pi$ | |
| | $\begin{bmatrix}-\sqrt{2}/2\\0\\\sqrt{2}/2\end{bmatrix} = \begin{bmatrix}-0.70711\\0\\0.70711\end{bmatrix}$ | $\pi$ | |

TABLE SII (continued).

| | | | |
|---|---|---|---|
| | $\begin{bmatrix}0\\1\\0\end{bmatrix}$ | $-\frac{\pi}{2},\frac{\pi}{2},\pi$ | |
| | $\begin{bmatrix}\frac{1}{2}\sqrt{2+\sqrt{2}}\\0\\\frac{1}{2}\sqrt{2-\sqrt{2}}\end{bmatrix}=\begin{bmatrix}0.92388\\0\\0.38268\end{bmatrix}$ | $\pi$ | |
| 2 | $\begin{bmatrix}-\frac{1}{2}\sqrt{2+\sqrt{2}}\\0\\\frac{1}{2}\sqrt{2-\sqrt{2}}\end{bmatrix}=\begin{bmatrix}-0.92388\\0\\0.38268\end{bmatrix}$ | $\pi$ | 7 |
| | $\begin{bmatrix}\frac{1}{2}\sqrt{2-\sqrt{2}}\\0\\\frac{1}{2}\sqrt{2+\sqrt{2}}\end{bmatrix}=\begin{bmatrix}0.38268\\0\\0.92388\end{bmatrix}$ | $\pi$ | |
| | $\begin{bmatrix}-\frac{1}{2}\sqrt{2-\sqrt{2}}\\0\\\frac{1}{2}\sqrt{2+\sqrt{2}}\end{bmatrix}=\begin{bmatrix}-0.38268\\0\\0.92388\end{bmatrix}$ | $\pi$ | |
| | $\begin{bmatrix}1\\0\\0\end{bmatrix}$ | $\pi$ | |
| 3 | $\begin{bmatrix}0\\1\\0\end{bmatrix}$ | $\pi$ | 3 |
| | $\begin{bmatrix}0\\0\\1\end{bmatrix}$ | $\pi$ | |
| | $\begin{bmatrix}0\\1\\0\end{bmatrix}$ | $\pi$ | |
| 4 | $\begin{bmatrix}\sqrt{2}/2\\0\\\sqrt{2}/2\end{bmatrix}=\begin{bmatrix}0.70711\\0\\0.70711\end{bmatrix}$ | $\pi$ | 3 |
| | $\begin{bmatrix}-\sqrt{2}/2\\0\\\sqrt{2}/2\end{bmatrix}=\begin{bmatrix}-0.70711\\0\\0.70711\end{bmatrix}$ | $\pi$ | |
| 5 | $\begin{bmatrix}0\\1\\0\end{bmatrix}$ | $\pi$ | 1 |
| 6 | None | None | 0 |

As an example, all system isometries for orientation 1 are visualized in Fig. S2. The original system is the identity, which is excluded from $n$ values. For the two-particle system in orientation 1, there are five unique axes of symmetry and a total of seven isometries. The degree of rotational symmetry for orientation 1 is therefore $n = 7$.

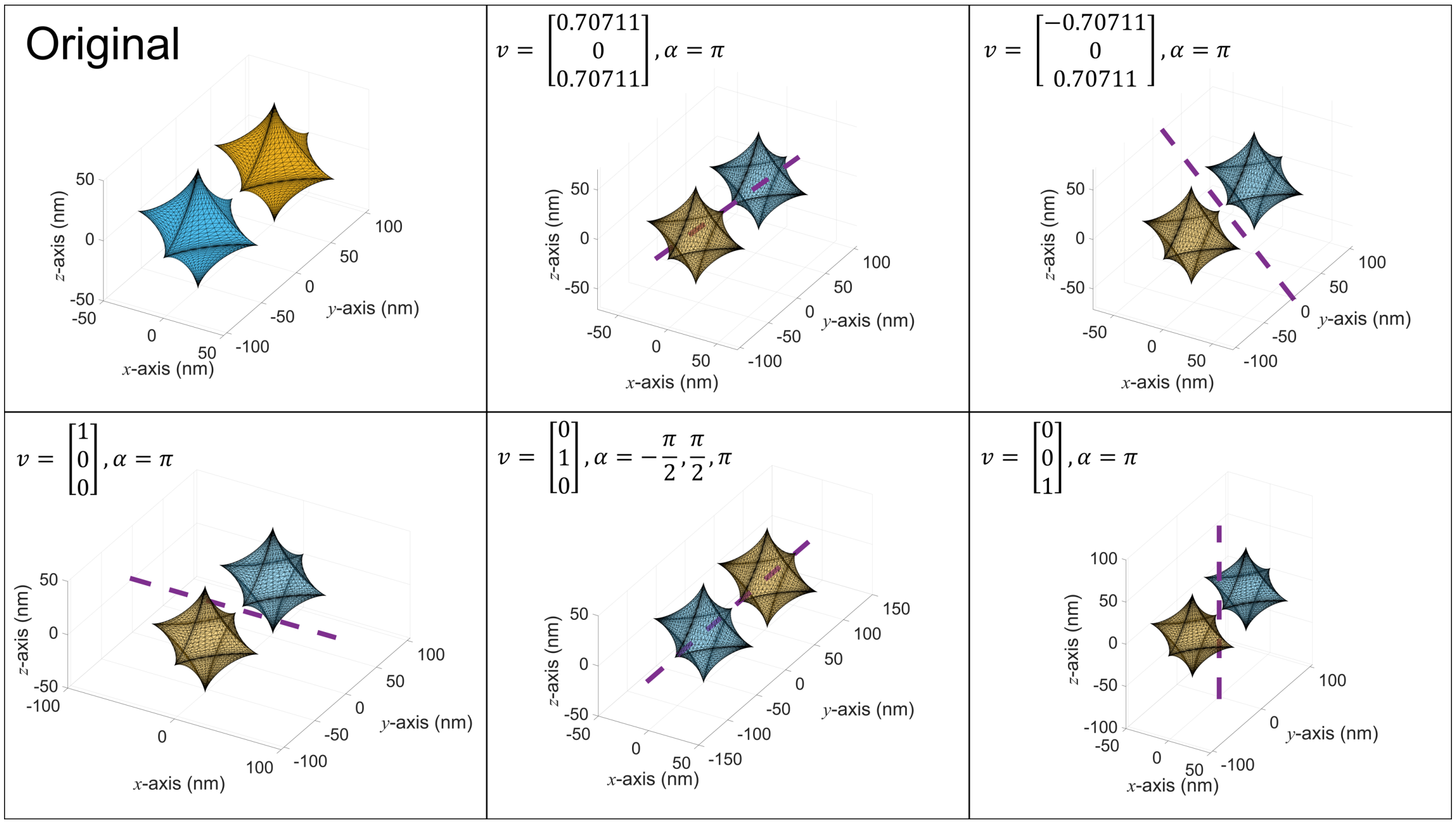

FIG. S2. All rotational isometries for orientation 1. The degree of rotational symmetry is $n = 7$ for this system. The particles shown are at a center-of-mass separation distance $d_c = 110$ nm, but the isometries remain valid for all center-of-mass separation distances. The coordinate system is located at the center of mass of the two-particle system.

## S3. COMPARISON OF THE MINIMUM VACUUM GAP DISTANCE BETWEEN TWO SUPERELLIPSOID $SiO_2$ PARTICLES FOR ALL ORIENTATIONS AND CENTER-OF-MASS SEPARATION DISTANCES

The minimum vacuum gap distance $d$ between each set of concave superellipsoid particles [Fig. S3(a)] and convex superellipsoid particles [Fig. S3(b)] for all orientations and center-of-mass separation distances $d_c$ is shown. Particles are of characteristic length $L_{ch} = 50$ nm.

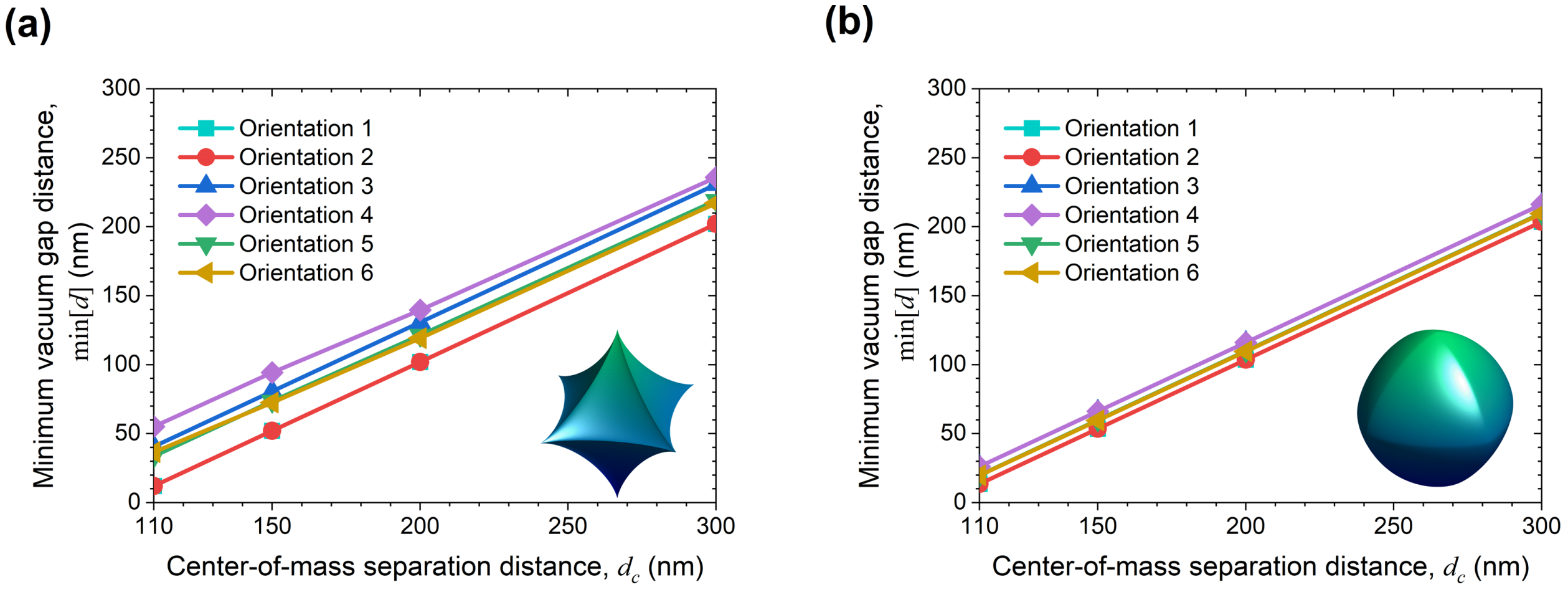

FIG. S3. Minimum vacuum gap distance versus center-of-mass separation distance for (a) concave superellipsoid and (b) convex superellipsoid $SiO_2$ particles for all six orientations.

## S4. COMPARISON OF TRENDS FOR THE NORMALIZED RANGE OF THE TOTAL CONDUCTANCE AND THE NORMALIZED RANGE OF THE MINIMUM VACUUM GAP DISTANCE FOR TWO SUPERELLIPSOID $SiO_2$ PARTICLES

The normalized range of the total conductance $\bar{R}(G_{t,AB})$ and the normalized range of the minimum vacuum gap distance $\bar{R}(\min[d])$ are compared for concave superellipsoid particles [Fig. S4(a)] and convex superellipsoid particles [Fig. S4(b)] at each center-of-mass separation distance $d_c$. Particles are of characteristic length $L_{ch} = 50$ nm. The normalized ranges $\bar{R}(G_{t,AB})$ and $\bar{R}(\min[d])$ display similar trends for variable particle separation distance $d_c$ and are strongly correlated.

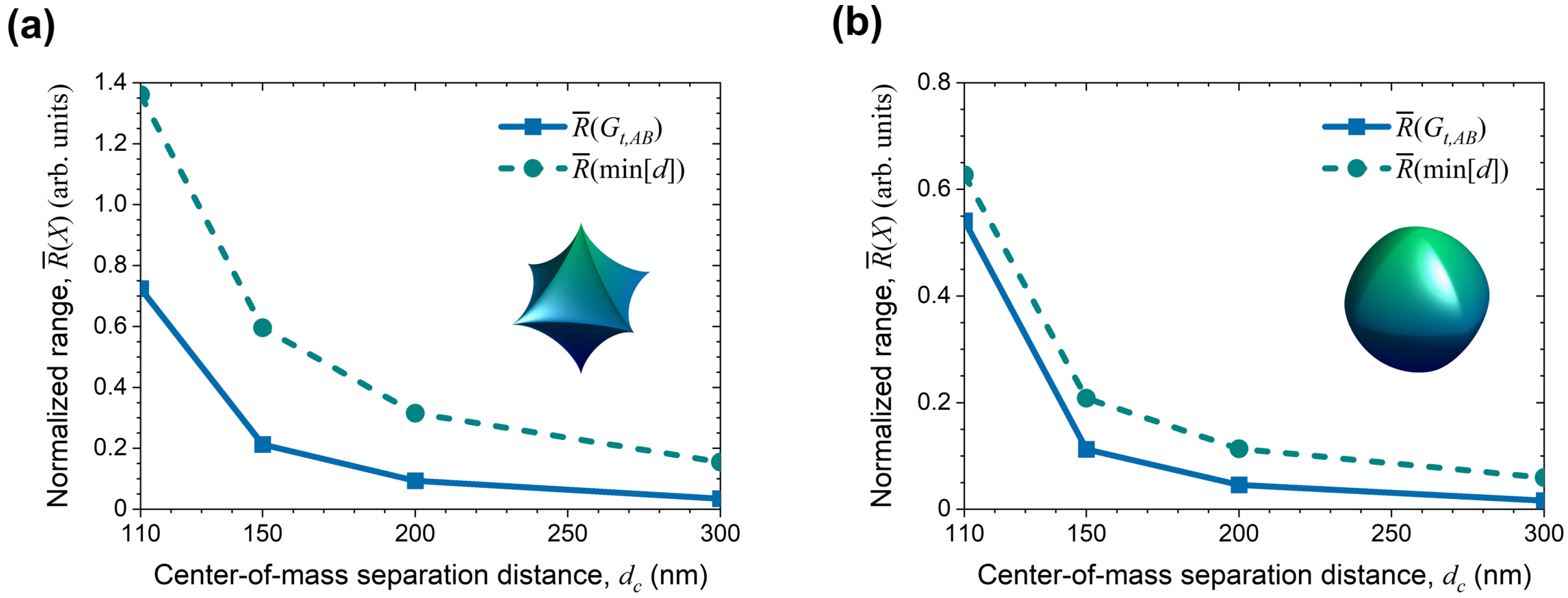

FIG. S4. Comparison of trends for the normalized range of the total conductance $\bar{R}(G_{t,AB})$ and the normalized range of the minimum vacuum gap distance $\bar{R}(\min[d])$ for (a) concave superellipsoid and (b) convex superellipsoid particles of variable center-of-mass separation distance. Particles are composed of $SiO_2$, and conductance is calculated at temperature $T = 300$ K.

## S5. SPECTRAL CONDUCTANCE BETWEEN TWO CONVEX SUPERELLIPSOID $SiO_2$ PARTICLES

The spectral conductance of convex superellipsoid particles [Figs. S5(a)-S5(d)] at all orientations and at each center-of-mass separation distance $d_c$ are compared. Particle orientation has the greatest influence on the total conductance at the closest center-of-mass separation distance $d_c = 110$ nm and is negligible at the farthest separation distance $d_c = 300$ nm.

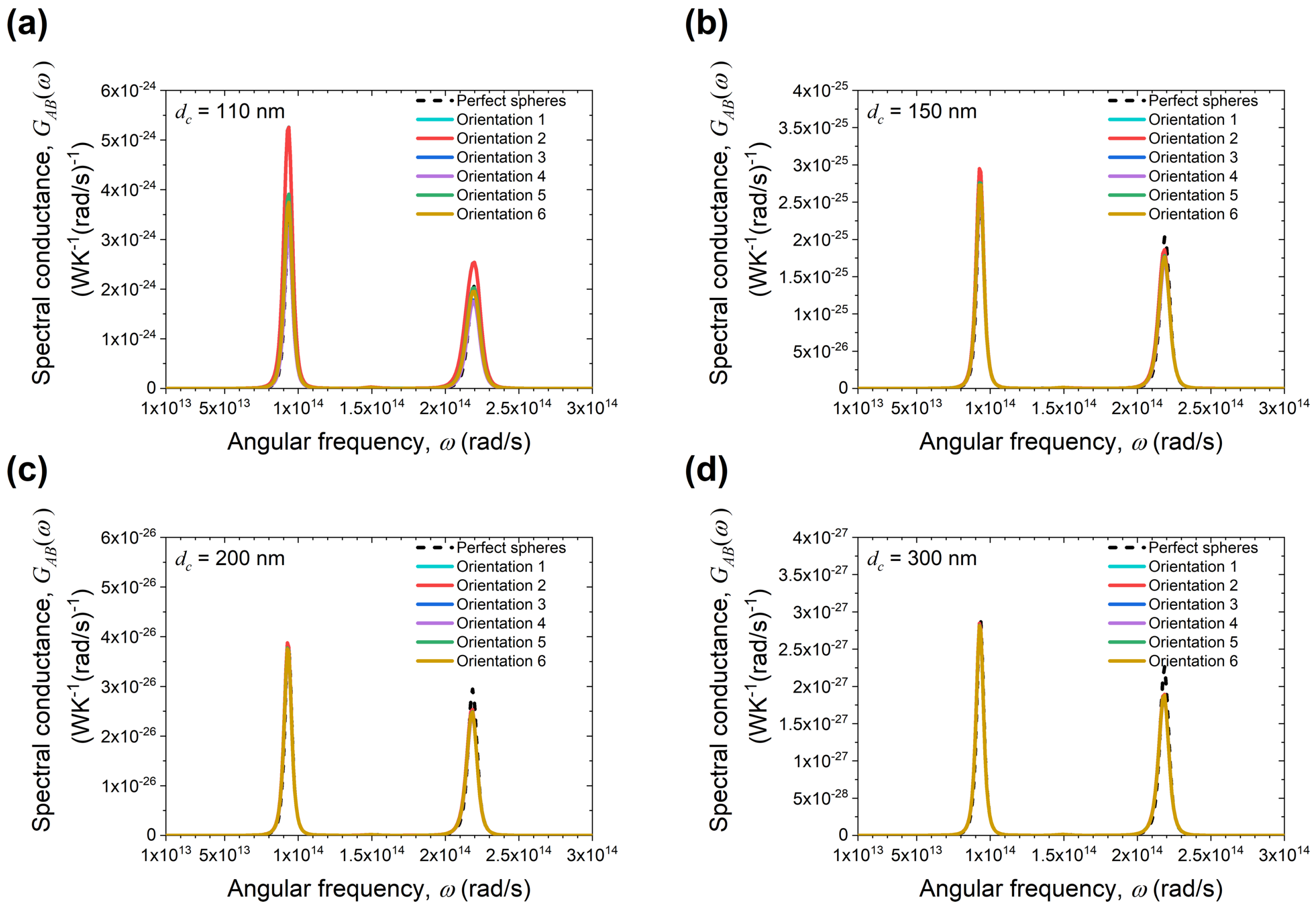

FIG. S5. Spectral conductance at temperature $T = 300$ K of convex superellipsoid $SiO_2$ particles at center-of-mass separation distances (a) $d_c = 110$ nm, (b) $d_c = 150$ nm, (c) $d_c = 200$ nm, and (d) $d_c = 300$ nm. The spectra of perfect spheres of equivalent volume and center-of-mass separation distance are calculated analytically using the method outlined in Ref. 3.